\documentclass[3p,times,twocolumn]{elsarticle}

\usepackage{amsmath,amssymb}
\usepackage[stable]{footmisc}

\newcommand{\beq}{\begin{equation}}
\newcommand{\eeq}{\end{equation}}
\newcommand{\nnnl}{\nonumber\\}
\renewcommand{\Im}{\text{Im}\,}
\renewcommand{\Re}{\text{Re}\,}
\newcommand{\Mpi}{M_\pi^2}
\newcommand{\M}{\mathcal{M}}
\newcommand{\F}{\mathcal{F}}
\newcommand{\mpi}{M_\pi}

\begin{document}

\begin{frontmatter}

\title{\vspace*{2.1cm}On the role of final-state interactions in Dalitz plot studies}

\author{Bastian Kubis}\ead{kubis@hiskp.uni-bonn.de}
\author{Franz Niecknig}
\author{Sebastian P.\ Schneider}

\address{Helmholtz-Institut f\"ur Strahlen- und Kernphysik and 
 Bethe Center for Theoretical Physics, Universit\"at Bonn, Germany}

\begin{abstract}
The study of Dalitz plots of heavy-meson decays to multi-hadron final states
has received intensified interest by the possibility to gain
access to precision investigations of CP violation.
A thorough understanding of the hadronic final-state interactions is a prerequisite 
to achieve a highly sensitive, model-independent study of such Dalitz plots. 
We illustrate some of the theoretical tools, predominantly 
taken from dispersion theory, available for these and related purposes,
and discuss the low-energy decays $\omega,\,\phi\to 3\pi$ in some more detail.
\end{abstract}


\end{frontmatter}

\section{Dalitz plots and CP violation\footnote{This section has significant
textual overlap with Ref.~\cite{hadron2011}.}}

A precise study of final-state interactions is increasingly becoming of high
importance for our understanding of the most diverse aspects of particle decays
involving hadrons.  
Final-state interactions can be of significance for various reasons: 
if they are strong, they can significantly enhance decay probabilities;
they can significantly \emph{shape} the decay probabilities, most prominently
through the occurrence of resonances; 
besides resonances, also new and non-trivial analytic structures can occur,
such as threshold or cusp effects (for the prominent role cusp effects have played
recently in studying pion--pion interactions, see Ref.~\cite{Cusp} and references therein); 
and finally, they introduce strong phases or imaginary parts, the existence of which 
is a prerequisite for the extraction of CP-violating phases 
in weak decays (see e.g.\ Ref.~\cite{BigiSanda}).
Dalitz plot studies of weak three-body decays of mesons with open heavy flavor (both $D$ and $B$)
are expected to acquire a key role in future precision investigations of CP violation,
due to their much richer kinematic freedom compared to the (effective) two-body 
final states predominantly used to study CP violation at the $B$ factories.  
In many cases, the branching fractions are significantly larger; furthermore,
the resonance-rich environment of multi-meson final states may help to enlarge small
CP phases in parts of the Dalitz plot, and differential observables may allow 
to obtain information on the operator structure that
drives CP violation beyond the Standard Model, once it is observed.
Since the results from the $B$ factories have shown that the Cabibbo--Kobayashi--Maskawa theory~\cite{CKM}
represents at least the dominant source of CP violation, our long-term goal will be to find other sources 
of CP violation that contribute additional, smaller effects. 
For this purpose, clearly extremely accurate measurements \emph{and} means of theoretical interpretation
are required.
Strong evidence for CP violation in three-body final states has already been reported
for $B^\pm \to K^\pm \pi^\mp \pi^\pm$~\cite{Belle_Babar-BCP}, e.g.\ with a $3.7\sigma$ signal
in the effective $K\rho$ channel. Only negative results exist 
for $D$ decays so far~\cite{Babar-DCP}; however with the first
preliminary evidence for direct CP violation in $D^0 \to K^+K^-$ vs.\ $D^0 \to \pi^+\pi^-$
reported~\cite{Charles-LHCb},
cross checks of any mechanism proposed for its explanation in three-body decays would be of
paramount importance.

There are different possibilities how to analyze CP violation in Dalitz plots.  
One suggestion is a strictly model-independent extraction from the data 
directly~\cite{susanmodelindep,miranda}, e.g.\ using the significance variable~\cite{miranda}
defined as
\beq
{}^{\rm Dp}S_{\rm CP}(i) \doteq \dfrac{N(i)-\bar N(i)}{\sqrt{N(i)+\bar N(i)}} ~,\eeq
where $N(i)$ and $\bar N(i)$ denote the event numbers of CP-conjugate decay modes
in a specific Dalitz plot bin $i$.
CP violation can then be identified in a deviation from a purely Gaussian distribution
in the significance plots.  The significance method allows to study local asymmetries
and requires no theoretical input at all.

An alternative approach is to make use of as much \emph{a priori} theoretical information 
on the strong amplitudes as possible.
To see why this may be advantageous, consider the following toy model:\footnote{I am grateful to 
C.~Hanhart for providing me with this example.}  let the event number distributions be
given by a (Breit--Wigner) resonance signal (of mass $M_{\rm res}$ and width $\Gamma_{\rm res}$, with strength 
$\beta$) on a certain background ($\propto \alpha$), with a CP-violating phase $\delta_{\rm CP}$, according to
\begin{align}
N, \, \bar N &= \alpha + \beta \,\Re \Bigg\{\frac{e^{\pm i \delta_{\rm CP}}}{s-M_{\rm res}^2+i M_{\rm res} \Gamma_{\rm res}}\Bigg\} \quad\Rightarrow\quad \nnnl
N-\bar N &= 
\frac{\sin\delta_{\rm CP} \times 2\beta M_{\rm res}\Gamma_{\rm res}}{(s-M_{\rm res}^2)^2+(M_{\rm res} \Gamma_{\rm res})^2} ~. \label{eq:demo_BW}
\end{align}
As the pole position of the resonance in question (given by $M_{\rm res}$ and $\Gamma_{\rm res}$)
is universal and may, for the sake of the argument, be accurately determined by other, 
independent processes, it is obvious that the \emph{functional form} of Eq.~\eqref{eq:demo_BW}
is strongly restricted.
Indeed, one can easily construct examples where the significance distribution 
shows no deviation from a Gaussian due to lack of data statistics~\cite{hadron2011}, 
while a fit with the strong amplitude~\eqref{eq:demo_BW} still allows 
to extract $\delta_{\rm CP}$ with some (limited) accuracy even in a sparse sample.
So while the theoretical assumptions and prejudices going into such an analysis clearly have to be
very carefully judged, their benefit in terms of vastly increased sensitivity is also obvious.

In the following, we will therefore briefly sketch some of the tools available,
mainly based on dispersion theory,
to analyze the hadronic amplitudes of the (light) final-state particles (such as pions and kaons):
(light) meson scattering, form factors for the description of two strongly interacting particles,
and the consistent treatment of three-body final states.

\section{Scattering and form factors}

Analyticity, unitarity, and crossing symmetry provide a high degree of constraint
for the pion--pion scattering amplitude. They can be exploited using dispersion relations,
which can be formulated as a coupled system of partial-wave equations, 
the so-called Roy equations~\cite{Roy}. 
Modern precision analyses of the Roy equations have been performed~\cite{ACGL+GarciaMartin},
partly making use of constraints from chiral perturbation theory
on the scattering lengths appearing as subtraction constants therein~\cite{CGL},
and a similarly rigorous study exists also for pion--kaon scattering~\cite{Buettiker}.
These provide us with high-precision parameterizations of the most relevant scattering amplitudes 
for light mesons appearing in the final states of heavy-meson decays.

\begin{figure}
  \begin{center}
    \includegraphics[width=\linewidth]{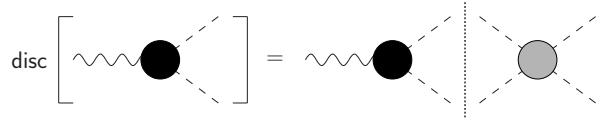} 
    \caption{Graphical representation of the consequence of analyticity and unitarity for form factors.}
    \label{fig:FFunit}
  \end{center}
\end{figure}

Final-state interactions between \emph{only two} strong\-ly interacting particles
can be described in terms of form factors, which in turn can be linked to the properties
of scattering amplitudes using analyticity and unitarity.  As illustrated in Fig.~\ref{fig:FFunit},
the unitarity relation for a form factor $F_J^I(s)$ (here: of the pion) of isospin $I$ and angular momentum $J$ reads
\beq
\Im F_J^I(s) = F_J^I(s) \times \theta \big(s-4\Mpi\big) \times \sin\delta_J^I(s) e^{-i\delta_J^I(s)} ~,
\label{eq:FF}
\eeq
from which one immediately deduces Watson's final-state theorem~\cite{Watson}: the form factor
shares the phase $\delta_J^I(s)$ of the (elastic) scattering amplitude.
The solution to Eq.~\eqref{eq:FF} is obtained in terms of the Omn\`es function~\cite{Omnes},
\begin{align}
F_J^I(s) &= P_J^I(s) \Omega_J^I(s) ~, \nnnl
\Omega_J^I(s)&=\exp\Bigg\{\frac{s}{\pi}\int_{4\Mpi}^\infty ds'\frac{\delta_J^I(s')}{s'(s'-s)} \Bigg\} ~,
\label{eq:Omnes}
\end{align}
where $P_J^I(s)$ is a polynomial. Note that the Omn\`es function is completely given in terms of the phase shift.
A classic application of such a form factor representation is the pion vector form factor 
$F_V^\pi(s)$, making use of the $\pi\pi$ P-wave phase $\delta_1^1(s)$;
for a precision parameterization up to about 1~GeV, 
$\rho-\omega$ mixing and the onset of (relatively week) inelasticities have 
to be accounted for~\cite{Troconiz,Bern:piFF}.
Such form factor representations can be used for analyses of the $e^+e^-\to\pi^+\pi^-$ data
to reduce the error of the hadronic contribution to the muon $g-2$, or to check the 
compatibility of the data with analyticity and unitarity~\cite{Bern:piFF}.
Note that the effects of chiral dynamics are particularly important for \emph{scalar} form 
factors, where a parameterization in terms of Breit--Wigner resonances 
can lead to completely wrong phase motions (see e.g.\ Ref.~\cite{susanandulf} for the context
of $B\to 3\pi$ decays).

\section{Dispersion relations for three-body decays: $\omega,\,\phi\to3\pi$}

\begin{figure}
  \begin{center}
    \includegraphics[width=0.5\linewidth]{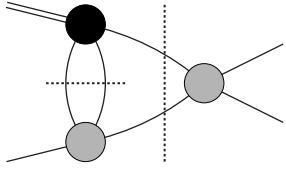} 
    \caption{Example for the complication of the analytic structure of 4-point functions 
             through crossed-channel effects, here for the decay of a heavy meson (double line)
             into three pions (single lines).}
    \label{fig:4Punit}
  \end{center}
\end{figure}

The application of dispersion relations to three-body decays is more complicated than 
the treatment of form factors due to the more involved analytic structure, and the possibility 
of crossed-channel rescattering; 
see Fig.~\ref{fig:4Punit} for a depiction of the complication
of the unitarity relation.  
One specific (low-energy) example 
that has received much renewed attention recently~\cite{eta3pi} 
is $\eta \to 3\pi$, due to its importance 
for the extraction of the light quark mass ratios.
Here, we concentrate instead on the even simpler three-pion decays of the lightest isoscalar
vector mesons, $\omega,\,\phi\to \pi^0\pi^+\pi^-$~\cite{Niecknig_inprogress}.
One starts by decomposing the amplitude $\M(s,t,u)$ according to
\beq
\M(s,t,u)=\epsilon_{\mu\nu\alpha\beta} n^\mu p_{\pi^+}^\nu p_{\pi^-}^\alpha p_{\pi^0}^\beta\,\F(s,t,u)~,
\eeq
where $n^\mu$ is the polarization vector of the decaying $\omega/\phi$, and 
$s=(p_{\pi^-}+p_{\pi^+})$, $t=(p_{\pi^-}+p_{\pi^0})^{2}$, $u=(p_{\pi^+}+p_{\pi^0})^2$.
Due to Bose symmetry, only partial waves of odd angular momentum contribute;
neglecting discontinuities of F- and higher partial waves, $\F(s,t,u)$ can be further decomposed as
\beq
\F(s,t,u)=\F(s)+\F(t)+\F(u)~.
\eeq 
The unitarity relation for $\F(s)$, assuming elastic final-state interactions, then leads 
to the following expression for the discontinuity of $\F(s)$:
\begin{align}
{\rm disc}\, \F(s) &= 2i\big\{\F(s) + \hat\F(s)\big\} \nnnl
&\times \theta \big(s-4\Mpi\big) \times \sin\delta_1^1(s) e^{-i\delta_1^1(s)} ~.
\label{eq:ImM}
\end{align}
Equation~\eqref{eq:ImM} is complicated compared to Eq.~\eqref{eq:FF}
by \emph{inhomogeneities} $\hat\F(s)$, which are given by angular averages over $\F$ according
to
\begin{align}
\hat\F(s) & =3\big\langle \big(1 - z^2\big)\F\big\rangle(s) ~, ~~
s_0 = \frac{1}{3}\big(M_V^2+3\mpi^2\big) ~,\nnnl
\big\langle z^nf\big\rangle(s) &
=\frac{1}{2}\int_{-1}^1 dz\,z^n f\Big(\tfrac{1}{2}\big(3s_0-s+z\kappa(s)\big)\Big) ~, \nnnl
\kappa(s)&=\lambda^{1/2}(M_V^2,\mpi^2,s) \sqrt{1-\frac{4\,\Mpi}{s}} ~,\label{eq:Mhat}
\end{align}
where $\lambda(x,y,z)=x^2+y^2+z^2-2(xy+xz+yz)$, and $M_V$ is the mass of the decaying vector meson.
Note that the angular integration including the $\kappa(s)$ function is non-trivial and generates
a complex analytic structure, including three-particle cuts due to the fact that $\omega$ and $\phi$ are 
unstable and decay.  
For details on how to properly deform the angular integration path, see e.g.\ Ref.~\cite{Bronzan}.
The analog to the Omn\`es solution~\eqref{eq:Omnes} are then integral equations involving
the inhomogeneities (compare also Ref.~\cite{AnisovichLeutwyler})
\beq
\F(s) = \Omega_1^1(s)\Biggl\{\alpha 
+\frac{s}{\pi}\int_{4\Mpi}^\infty\frac{ds'}{s'}\frac{\sin\delta_1^1(s')\hat\F(s')}{|\Omega_1^1(s')|(s'-s)}\Biggr\} ~, \label{eq:inhomOmnes}
\eeq
with the subtraction constant $\alpha$.  
The number of subtractions is chosen such that the dispersion integral is guaranteed to converge.
(See Ref.~\cite{KhuriTreiman+Aitchison} for earlier, related
formulations.) 
Again, care has to be taken when performing the dispersive integral, as $\hat\F(s)$ shows singular behavior
at the pseudothreshold $s=(M_V-\mpi)^2$.

Equations~\eqref{eq:Mhat} and \eqref{eq:inhomOmnes} can be solved iteratively:
starting from an arbitrary input function $\F(s)$, we can calculate the 
inhomogeneity $\hat\F(s)$ according to Eq.~\eqref{eq:Mhat}, from which a new $\F(s)$
is obtained from Eq.~\eqref{eq:inhomOmnes}, provided we can devise a method to determine
the subtraction constant; the procedure is stopped once a fixed point of the iteration
is reached with sufficient accuracy.
In the example discussed here, see Eq.~\eqref{eq:inhomOmnes}, 
the subtraction constant works as an overall normalization factor
of the solution; we match it to the partial decay width, but note that a \emph{normalized} Dalitz plot 
distribution is subsequently a pure prediction.
While the result is independent of the starting function, for the case at hand, we choose
$\F(s) = \Omega_1^1(s)$ in order to allow us to quantify crossed-channel effects (generated
by the iteration) in a plausible way.

\begin{figure}[t!]
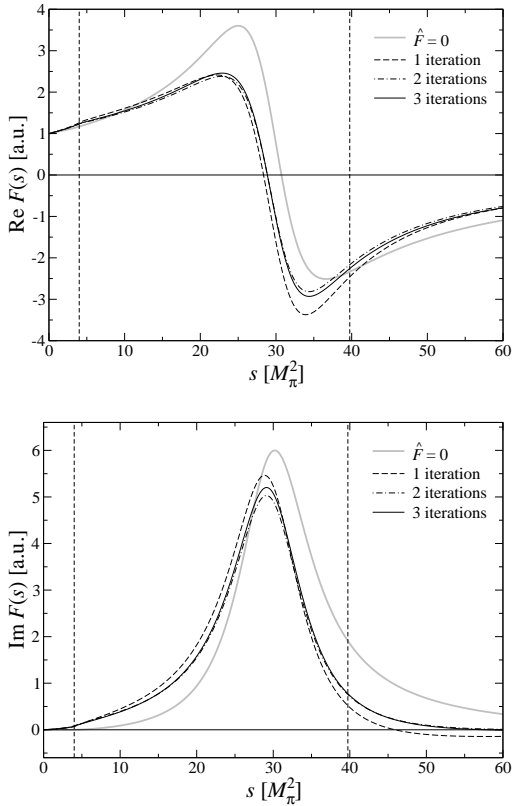

 \centering 
\includegraphics[width= 0.85\linewidth]{iterationphi_Re} \\[4mm]
\includegraphics[width= 0.85\linewidth]{iterationphi_Im}
 \caption{Successive iteration steps of real (upper panel) and imaginary (lower panel) 
part of the amplitude $\F(s)$ for
$\phi\to 3\pi$. The dashed lines denote the physical region of the decay.}
\label{fig:amplitudephi}
\end{figure}

\begin{figure}
\includegraphics*[width = \linewidth]{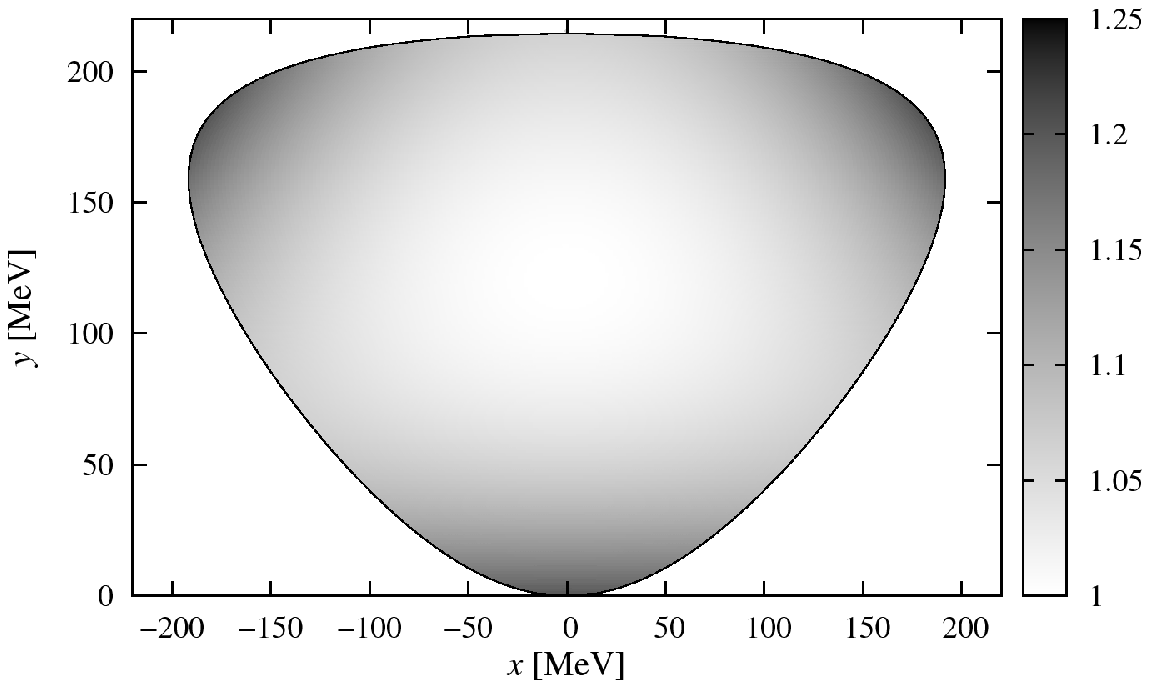} \\[4mm]
\includegraphics*[width = \linewidth]{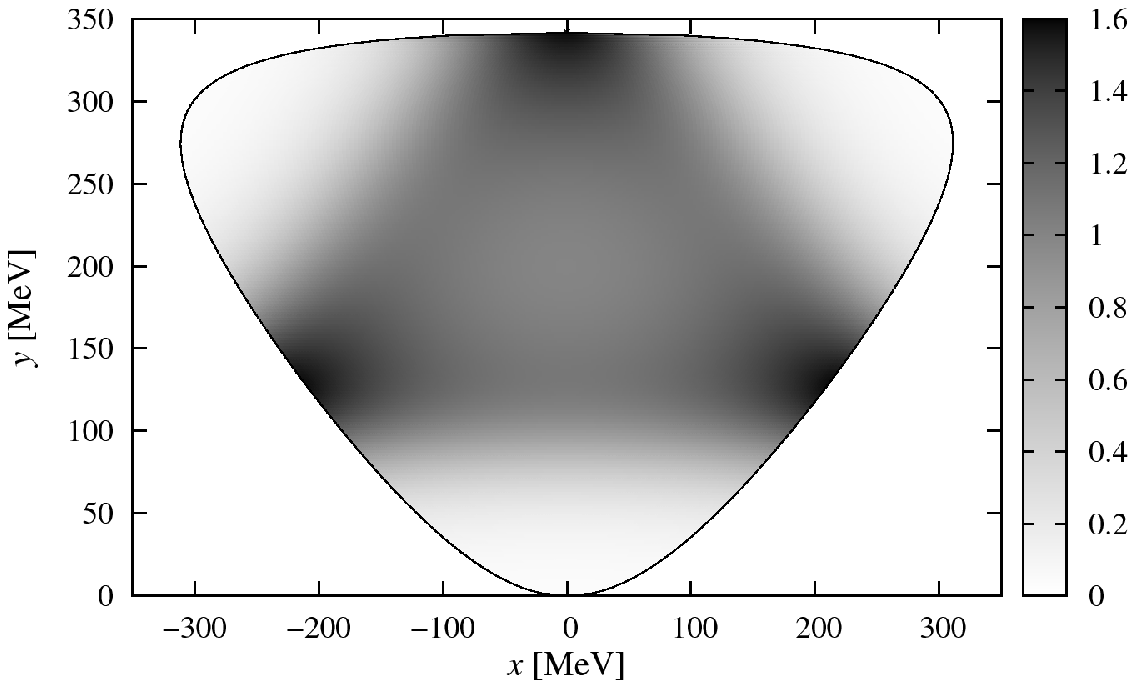}
\caption{Dalitz plots for $\omega\to3\pi$ (top) and $\phi\to3\pi$ (bottom), normalized
by the P-wave phase space factor.}
\label{fig:Dalitz}
\end{figure}

Figure~\ref{fig:amplitudephi} shows the result of such an iteration for the decay $\phi\to 3\pi$:  
it converges fast, with the third iteration already all but indistinguishable from the final result.
The difference to the starting point of the iteration, the Omn\`es function without any
crossed-channel rescattering, is however very significant.
The picture for $\omega\to 3\pi$ (not shown here) is qualitatively very similar, with 
convergence reached even faster (after two iterations).
The resulting Dalitz plots for both $\omega\to 3\pi$ and $\phi\to 3\pi$ are shown in
Fig.~\ref{fig:Dalitz}, normalized by the  P-wave phase space factor
(such that what is effectively shown is $|\F(s,t,u)|^2$ in arbitrary normalization),
using the kinematical variables
\beq
x=\frac{t-u}{2M_V} ~, \quad y =\frac{(M_V-\mpi)^2-s}{2M_V} ~.
\eeq
The $\omega\to 3\pi$ Dalitz plot distribution is relatively smooth, 
rising from the center to its borders, with a maximum enhancement in the outer corners
of about 20\%.
The $\phi\to 3\pi$ one displays significantly more structure, as the $\rho$ resonance
bands lie inside the physical decay region: the Dalitz plot distribution 
again rises from its center to the three $\rho$ bands, before falling off steeply 
towards the corners.
The profile of the  $\phi\to 3\pi$ Dalitz plot agrees qualitatively with 
experimental results~\cite{phi3piexp};
a detailed comparison is in progress~\cite{Niecknig_inprogress}.

\begin{figure}[t!]
\includegraphics*[width = \linewidth]{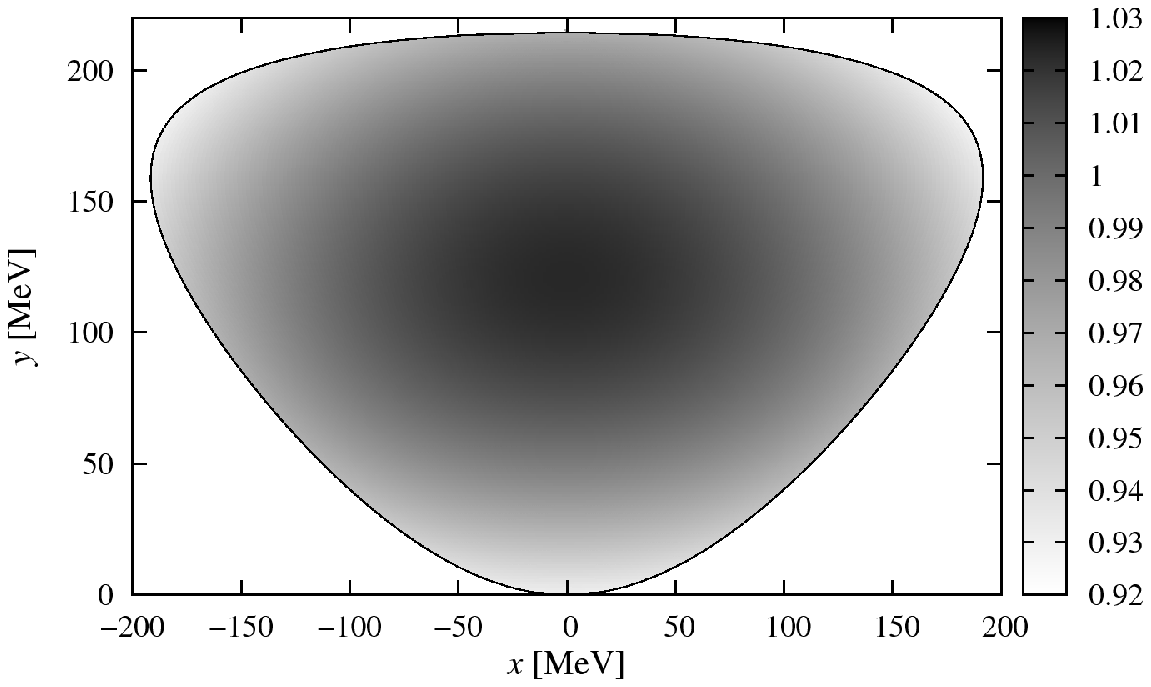} \\[4mm]
\includegraphics*[width = \linewidth]{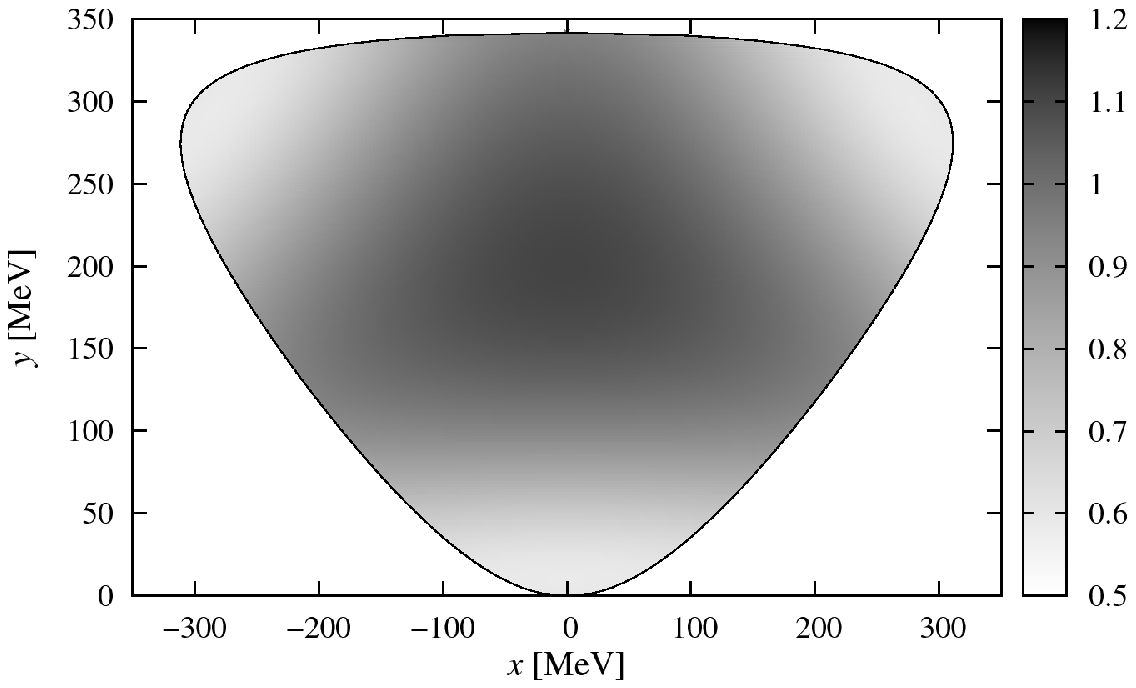}
\caption{$|\F_{\rm full}|^2/|\F_{\hat\F=0}|^2$ for $\omega\to3\pi$ (top) and $\phi\to3\pi$ (bottom). The subtraction constant is fixed so that it reproduces the decay width before and after the iteration.}
\label{fig:DalitzfixedGamma}
\end{figure}

In order to illustrate the effects of crossed-channel rescattering explicitly, we compare the full
dispersive solution to the approximation with inhomogeneities $\hat F(s)$ in Eq.~\eqref{eq:inhomOmnes}
neglected.
By keeping the subtraction constant $\alpha$ fixed (assuming, say, it can be determined using 
some other theoretical insight), we find that crossed-channel effects enhance the $\omega\to 3\pi$
partial width by about 20\%, while the $\phi\to 3\pi$ partial width, in contrast, is \emph{reduced}
by about the same amount.  As far as the Dalitz plots are concerned, this effect
can however largely be absorbed in a redefinition of the subtraction constant.
The remaining effect on the normalized distributions is demonstrated in Fig.~\ref{fig:DalitzfixedGamma}:
crossed-channel effects increase the Dalitz plot densities in the center, and decrease them towards
the borders, where the size of the effects is significantly larger for the decay of the $\phi$;
the $\rho$ resonance bands  in the case of $\phi\to 3\pi$ are left rather untouched.

The method sketched here is currently also applied to other light-meson decays such as 
$\eta'\to\eta\pi\pi$~\cite{Schneider_inprogress}. Challenges to be faced when 
extending this formalism to heavy-meson decays include the necessity to treat systems of
integral equations when coupled channels within one partial wave cannot be ignored, or
inelasticities are not negligible.  
In particular when considering $B$-meson decays, elastic unitarity will surely not be sufficient.
It will have to be checked when a purely perturbative
treatment of crossed-channel effects is feasible (compare e.g.\ Ref.~\cite{Liu}), and when 
higher partial waves become important.
To investigate these and related questions is part of the program of the informal
{\it Les Nabis} network~\cite{LesNabis}, which 
aims at optimizing future Dalitz plot studies along the lines sketched here---with the 
strong goal to better interpret the mechanism of CP violation in nature, yet at the same time
learning important lessons on nonperturbative strong interactions.

\section*{Acknowledgements}

B.K.\ would like to thank the organizers of PhiPsi11 for inviting him to such a wonderful workshop,
and for their warm hospitality.
We are grateful to C.~Hanhart for stimulating discussions concerning some of the material presented here.
Partial financial support by
by DFG (SFB/TR 16, ``Subnuclear Structure of Matter''),
by the project ``Study of Strongly Interacting Matter'' 
(HadronPhysics2, grant 227431) under the 7th Framework Program of the EU,
and by the Bonn--Cologne Graduate School of Physics and Astronomy
is gratefully acknowledged.

\bibliographystyle{elsarticle-num}

\begin{thebibliography}{00}

\bibitem{hadron2011}
  B.~Kubis,
  arXiv:1108.5866 [hep-ph].

\bibitem{Cusp}
  J.~Gasser, B.~Kubis and A.~Rusetsky,
  Nucl.\ Phys.\  B {\bf 850} (2011) 96
  [arXiv:1103.4273 [hep-ph]];
%
  B.~Kubis,
  EPJ Web Conf.\  {\bf 3} (2010) 01008
  [arXiv:0912.3440 [hep-ph]].

\bibitem{BigiSanda}
  I.~I.~Bigi and A.~I.~Sanda,
  {\em CP violation} (2nd ed.),
  Camb.\ Monogr.\ Part.\ Phys.\ Nucl.\ Phys.\ Cosmol.\  {\bf 28} (2009) 1.

\bibitem{CKM}
  N.~Cabibbo,
  Phys.\ Rev.\ Lett.\  {\bf 10} (1963) 531;
%
  M.~Kobayashi and T.~Maskawa,
  Prog.\ Theor.\ Phys.\  {\bf 49} (1973) 652.

\bibitem{Belle_Babar-BCP}
  A.~Garmash {\it et al.}  [Belle Collaboration],
  Phys.\ Rev.\ Lett.\  {\bf 96} (2006) 251803
  [arXiv:hep-ex/0512066];
%
  B.~Aubert {\it et al.}  [BABAR Collaboration],
  Phys.\ Rev.\  D {\bf 78} (2008) 012004
  [arXiv:0803.4451 [hep-ex]].

\bibitem{Babar-DCP}
  B.~Aubert {\it et al.} [BABAR Collaboration],
  Phys.\ Rev.\  D {\bf 78} (2008) 051102
  [arXiv:0802.4035 [hep-ex]].

\bibitem{Charles-LHCb}
  M.~Charles [LHCb Collaboration], talk presented at HCP2011
  (Nov.\ 14--18, 2011, Paris, France), {\tt http:// hcp2011.lpnhe.in2p3.fr},
  LHCb-CONF-2011-061.

\bibitem{susanmodelindep}
  S.~Gardner,
  Phys.\ Lett.\  B {\bf 553} (2003) 261
  [arXiv:hep-ph/0203152];
%
  S.~Gardner and J.~Tandean,
  Phys.\ Rev.\  D {\bf 69} (2004) 034011 
  [arXiv:hep-ph/0308228].

\bibitem{miranda}
  I.~Bediaga, I.~I.~Bigi, A.~Gomes, G.~Guerrer, J.~Miranda and A.~C.~d.~Reis,
  Phys.\ Rev.\  D {\bf 80} (2009) 096006
  [arXiv:0905.4233 [hep-ph]].

\bibitem{Roy}
  S.~M.~Roy,
  Phys.\ Lett.\  B {\bf 36} (1971) 353.

\bibitem{ACGL+GarciaMartin}
  B.~Ananthanarayan, G.~Colangelo, J.~Gasser and H.~Leutwyler,
  Phys.\ Rept.\  {\bf 353} (2001) 207
  [arXiv:hep-ph/0005297];
%
  R.~Garc\'ia-Mart\'in, R.~Kami\'nski, J.~R.~Pel\'aez, J.~Ruiz~de~Elvira and F.~J.~Yndur\'ain,
  Phys.\ Rev.\  D {\bf 83} (2011) 074004
  [arXiv:1102.2183 [hep-ph]].

\bibitem{CGL}
  G.~Colangelo, J.~Gasser and H.~Leutwyler,
  Nucl.\ Phys.\  B {\bf 603} (2001) 125
  [arXiv:hep-ph/0103088].

\bibitem{Buettiker}
  P.~B\"uttiker, S.~Descotes-Genon and B.~Moussallam,
  Eur.\ Phys.\ J.\  C {\bf 33} (2004) 409
  [arXiv:hep-ph/0310283]

\bibitem{Watson}
  K.~M.~Watson,
  Phys.\ Rev.\  {\bf 95} (1954) 228.

\bibitem{Omnes}
  R.~Omn\`es,
  Nuovo Cim.\  {\bf 8} (1958) 316.

\bibitem{Troconiz}
  J.~F.~De Troc\'oniz and F.~J.~Yndur\'ain,
  Phys.\ Rev.\  D {\bf 65} (2002) 093001
  [arXiv:hep-ph/0106025].

\bibitem{Bern:piFF}
  H.~Leutwyler,
  arXiv:hep-ph/0212324;
%
  G.~Colangelo,
  Nucl.\ Phys.\ Proc.\ Suppl.\  {\bf 131} (2004) 185
  [arXiv:hep-ph/0312017].

\bibitem{susanandulf}
  S.~Gardner and U.-G.~Mei{\ss}ner,
  Phys.\ Rev.\  D {\bf 65} (2002) 094004
  [arXiv:hep-ph/0112281].

\bibitem{eta3pi}
  G.~Colangelo, S.~Lanz and E.~Passemar,
  PoS {\bf CD09} (2009) 047
  [arXiv:0910.0765 [hep-ph]];
%
  S.~P.~Schneider, B.~Kubis and C.~Ditsche,
  JHEP {\bf 1102} (2011) 028
  [arXiv:1010.3946 [hep-ph]];
%
  K.~Kampf, M.~Knecht, J.~Novotn\'y and M.~Zdr\'ahal,
  arXiv:1103.0982 [hep-ph].

\bibitem{Niecknig_inprogress}
  F.~Niecknig, B.~Kubis and S.~P.~Schneider, {\it work in progress}.

\bibitem{Bronzan} 
  J.~B.~Bronzan and C.~Kacser,
  Phys.\ Rev.\ {\bf 132} (1963) 2703.


\bibitem{AnisovichLeutwyler}
  A.~V.~Anisovich and H.~Leutwyler,
  Phys.\ Lett.\  B {\bf 375} (1996) 335
  [arXiv:hep-ph/9601237].

\bibitem{KhuriTreiman+Aitchison}
  N.~N.~Khuri and S.~B.~Treiman,
  Phys.\ Rev.\  {\bf 119} (1960) 1115;
%
  I.~J.~R.~Aitchison,
  J.\ Phys.\ G {\bf 3} (1977) 121.

\bibitem{phi3piexp}
  R.~R.~Akhmetshin {\it et al.},
  Phys.\ Lett.\  B {\bf 642} (2006) 203;
%
  A.~Aloisio {\it et al.}  [KLOE Collab.],
  Phys.\ Lett.\  B {\bf 561} (2003) 55
  [Erratum-ibid.\  B {\bf 609} (2005) 449]
  [arXiv:hep-ex/0303016].

\bibitem{Schneider_inprogress}
  S.~P.~Schneider and B.~Kubis, {\it work in progress}.

\bibitem{Liu}
  B.~Liu, M.~B\"uscher, F.~K.~Guo, C.~Hanhart and U.-G.~Mei{\ss}ner,
  Eur.\ Phys.\ J.\  C {\bf 63} (2009) 93 
  [arXiv:0901.1185 [hep-ph]].

\bibitem{LesNabis}
  I.~I.~Bigi, S.~Gardner, C.~Hanhart, B.~Kubis, T.~Mannel, U.-G.~Mei\ss ner, W.~Ochs, J.~A.~Oller, 
  J.R.~Pel\'aez, M.R.~Pennington, A.~Sibirtsev ({\it theory});
  I.~Bediaga, A.E.~Bondar, A.~Denig, T.~J.~Gershon, W.~Gradl, B.~T.~Meadows, K.~Peters, U.~Wiedner, G.~Wilkinson
  ({\it experiment})
  {\it et al.} [Les Nabis Collaboration].

\end{thebibliography}

\end{document}